\documentstyle[11pt,newpasp,twoside]{article}
\def\edcomment#1{\iffalse\marginpar{\raggedright\sl#1\/}\else\relax\fi}
\reversemarginpar

\def\lsim{{{}_{{}_<}^{~}\atop {}^{{}^\sim}_{~}}}
\newcommand{\be}{\begin{equation}}
\newcommand{\ee}{\end{equation}}

\begin{document}
\title{The Distance to the Large Magellanic Cloud}
 \author{Piotr Popowski}
\affil{Institute of Geophysics and Planetary Physics, L-413\\ Lawrence Livermore
National Laboratory, University of California\\ P. O. Box 808, Livermore, CA 94551, USA }

\begin{abstract}
I demonstrate that the two unexpected results in the local Universe: 
anomalous intrinsic
$(V-I)_0$ colors of RR Lyrae stars and clump giants in the Galactic center, and
very short distances to Magellanic Clouds inferred from clump giants, can be
at least partially resolved with a modified coefficient of selective extinction
$A_V/E(V-I)$.
With this modification, I find a new clump-giant distance modulus to
the Large Magellanic Cloud, $\mu_{\rm LMC} = 18.27 \pm 0.07$, which is 0.09 
larger than
the Udalski (1998b) result. When distance estimates from the red clump,
RR Lyrae stars and the eclipsing binary HV2274 are combined, one obtains
$\mu_{\rm LMC} = 18.31 \pm 0.04 \,\,\,{\rm (internal)}$.
\end{abstract}

\section{Distance to the LMC -- Controversy and New Determinations}

The Hubble constant, $H_0$, is one of the most important cosmological
parameters.
There are two major paths to determine $H_0$.
The more elegant one is non-local and based on observations of the 
high-redshift Universe. Modeling of gravitational lensing of quasars and observations of
the Cosmic Microwave Background both belong to this group of methods.
The second path goes through a determination of distances and recession
velocities of objects with motions dominated by the Hubble linear expansion.
The distances, which are harder to measure than recession velocities,
are determined based on a distance-ladder approach. 
The Magellanic Clouds, and especially the Large Magellanic Cloud (LMC),
play a major role in this treatment. Almost all of the extragalactic distance
scale is known only relative to the LMC (Madore et al. 1999). Therefore, it is 
absolutely essential to establish a reliable distance to the LMC 
($d_{\rm LMC}$). 
This can be achieved only with an understanding of the systematics inherent
in the standard candles used for the distance determinations.

For many years now there has been a division between the so called ``short''
and ``long'' distance scales to the LMC. Currently, the measured values of
$d_{\rm LMC}$ span a range of over 25\%  (see e.g., Feast \& 
Catchpole 1997; Stanek, Zaritsky, \& Harris 1998).

Distances are measured based on a general formula
\be
\mu_{\rm LMC} = X - M_X - A_X \label{mu}
\ee
where $X$ is an apparent magnitude, $M_X$ is an absolute magnitude, and $A_X$ 
is an extinction in the band of the observations. The uncertainties in all
ingredients present in equation (1) may compromise the final answer.
Here I am going to concentrate on three new or revised methods that
cluster consistently around a short distance to the LMC:\\[0.1cm]
[1] Paczy\'{n}ski \& Stanek (1998) pointed out that clump giants 
should constitute an accurate
distance indicator. Udalski et al. (1998b) and Stanek et al. (1998) applied 
the clump method and found a very short distance to the LMC ($\mu_{\rm LMC} 
\lsim 
18.2$). In response, Cole (1998) and Girardi et al. (1998) suggested that 
clump giants are not standard candles and that their
absolute $I$ magnitudes, $M_I({\rm RC})$, depend on the metallicity and age 
of the
population. Udalski (1998a, 1998b) rejected this criticism by showing
that the metallicity dependence is
at a low level of about $0.1$ mag/dex, and that $M_I({\rm RC})$ is 
approximately constant for cluster ages between 2 and 10 Gyr.
Recent developments (see e.g., Popowski 2000 for
a short review) suggest that the absolute character of $M_I({\rm RC})$ 
is a major systematic uncertainty in this method.\\[0.08cm]
[2] Popowski \& Gould (1999) determined the absolute magnitude of 
RR Lyrae stars, $M_V({\rm RR}) = 0.71 \pm 0.07$ at ${\rm
[Fe/H]} = -1.6$, from the statistical parallax, cluster kinematics and 
trigonometric parallax methods. When this result is coupled with the LMC RR 
Lyrae photometry of Udalski et al. (1999) and Walker (1992), one obtains
$\mu_{\rm LMC} \approx 18.30 \pm 0.08$.
The value of $M_V({\rm RR})$ remains the main uncertainty of this 
determination.\\[0.08cm]
[3] Guinan et al. (1998) solved the eclipsing binary HV2274 and obtained
various stellar parameters and the distance to the LMC. The spectra used
for this purpose did not extend far enough toward long wavelengths, and the 
$B$ and $V$ photometry was needed to break the 
degeneracy
between the reddening and the shape of the extinction curve. 
With Udalski et al. (1998c) photometry,  Guinan et al. (1998) obtained 
$\mu_{\rm LMC} = 18.30 \pm 0.07$. Application of Nelson et al. (2000) 
photometry would result in $\mu_{\rm LMC} = 
18.40 \pm 0.07$.  The reddening constitutes a major 
uncertainty. \\[-0.20cm]

In two out of three cases the absolute magnitudes of distance indicators
are under debate. 
Due to a huge number of possible environments, it is very hard to
prove the standard character of a given candle. However, it should be
possible to check whether other stellar characteristics of a candle  behave in
a predictable fashion. To follow this suggestion, I will concentrate on the
stars in the Galactic bulge.

\section{The Mystery of Anomalous Colors in the Galactic Bulge}

Paczy\'{n}ski (1998) tried to explain
why the clump giants
in the Baade's Window have $(V-I)_0$ colors which are approximately $0.2$
mag redder than in the solar neighborhood. He could not find any
satisfactory answer. Stutz, Popowski, \& Gould (1999) found a corresponding
effect for the Baade's Window RR Lyrae stars, which have $(V-I)_0$ redder by
about 0.17 than their local counterparts.
The similar size of the color shift in RR Lyrae stars and clump giants
suggests common origin. 
The bulge RR Lyrae stars and clump giants both 
burn Helium in their cores, but similarities end there. 
RR Lyrae stars pulsate, clump giants do not. RR Lyrae stars are metal-poor,
clump giants are metal-rich.
RR Lyrae stars are a part of an axisymmetric stellar halo, whereas clump 
giants form a bar.
Stutz et al. (1999) suggested that the very red $(V-I)_0$ of the bulge 
RR~Lyrae
stars might have resulted from an unusual abundance of $\alpha$ elements.
Why should a clump population which emerged in a different formation
process share the same property?

The presence of the same type of color anomaly
for different types of stars suggests that the effect might be unrelated
to the physics of those stars.
The investigated RR Lyrae and clump giants share
two things in common.
 First, photometry of both types of stars comes
from the OGLE, phase-I, project. Indeed, Paczy\'{n}ski et
al. (1999) showed that
OGLE-I $V$-magnitudes are 0.021 mag fainter, and $I$-magnitudes 0.035 mag brighter than better calibrated OGLE-II magnitudes. Therefore, the correct $(V-I)$ colors
should be 0.056 bluer. Additionally, the new $(V-I)_0$ from the more 
homogeneous
Baade's Window clump is bluer than Paczy\'{n}ski's \& Stanek's (1998) color,
even when reduced to OGLE-I calibration. As a result, the
$(V-I)_0$ anomaly shrinks and the remaining shift amounts to $\sim 0.11$ for both the RR Lyrae stars and clump giants.
Second, Paczy\'{n}ski (1998) and Stutz et al. (1999) use the same
extinction map (Stanek 1996) and the same coefficient of conversion from 
visual 
extinction $A_V$ to a color excess $E(V-I)$. The absolute values of the $A_V$s 
are probably correct, because the zero-point of the extinction
map was determined 
from the $(V-K)$ color, and $A_V/E(V-K)$ is very close to 1.
However, $R_{VI}=A_V/E(V-I)$ is not as secure, and has a pronounced effect
on the obtained color.

The value and variation of $R_{VI}$ was thoroughly investigated by
Wo\'{z}niak \& Stanek (1996).
The essence of the Wo\'{z}niak \& Stanek (1996) method to determine 
differential extinction is 
an assumption that regions of the sky with a lower surface
density of stars have higher extinction.
 This is quite a natural 
expectation, as far as the density of the underlying true population of stars does not depend on location.
However, it is not obvious a priori how to convert a certain density of stars
to an amount of visual extinction. Therefore, Wo\'{z}niak \& Stanek (1996) used
clump giants to calibrate their extinction. To make a calibration procedure
completely unbiased would require, among other things, that the $V$-magnitudes
of clump giants do not depend on their color [here $(V-I)_0$], that
reddened and unreddened clump giants be drawn from the same parent population,
and that clump giants were selected without any assumption about $R_{VI}$.
None of those is true (for details see Popowski 2000).

Because the smaller selective extinction coefficient is not excluded by 
the 
current studies, I will assume $R_{VI}=2.1$ to match the $(V-I)_0$ colors 
of 
the bulge with the ones in the solar neighborhood. The color is a weak 
function of [Fe/H], so this procedure is justified because the 
metallicities 
of the bulge and solar neighborhood are similar. The change in $R_{VI}$ 
from
2.5 to 2.1 will decrease the $I$-mag extinction by 0.11 mag, and increase 
the 
clump-based distance to the Galactic center by the same amount.

%We will match 
%$(V-I)_0$ 
%colors of the bulge with the ones in the solar neighborhood, because
%color should be a function of metallicity [Fe/H], and [Fe/H] of the bulge
%and solar neighborhood are very similar.
%Because it is conceivable that the selective extinction
%coefficient could be smaller, we modify $R_{VI}$ in the bulge from 2.5 
%to 2.1 to achieve a change of anomalous Baade's Window colors of
%$\Delta (V-I)_0 \approx -0.11$. Consequently, the $I$-band extinction 
%and the 
%clump-based distance to the Galactic center are influenced as well.

\section{Recalibration of Clump Giant Stars}

How do the bulge results bear on the distance to the LMC?
The better photometry from Paczy\'{n}ski et al. (1999) and the modification
of $R_{VI}$ influence the relative RR Lyrae and clump distances to 
the Galactic center. Thus, the $M_I({\rm RC})$ -- [Fe/H] relation for clump 
giants used in the LMC, which was calibrated with respect to the baseline 
provided by RR Lyrae stars (Udalski 1998a), changes.
Assuming linearity of $M_I({\rm RC})$ -- [Fe/H] 
and making some small adjustments to [Fe/H] used
by Udalski (1998a), I find:
\be
M_I({\rm RC}) = -0.23 + 0.19 {\rm [Fe/H]}, \label{mi}
\ee
with a slope 0.10 mag/dex steeper than the original result.
Such adjustment increases the best clump giant estimate from 
$\mu_{\rm LMC} = 18.18 \pm 0.06$ to $\mu_{\rm LMC}= 18.27 \pm 0.07$.
When distance estimates from the red clump, RR Lyrae stars and eclipsing 
binary HV2274 are combined, one obtains $\mu_{\rm LMC} = 18.31 \pm 0.04$. 
The $1 \sigma$ uncertainty of this determination is only a formal error. 
The systematic errors are likely to dominate the true uncertainty. 
However, if the methods presented do not suffer from severe biases, 
then a distance to the LMC as long as 52-55 kpc 
($\mu_{\rm LMC} \approx$ 18.6-18.7) is highly disfavored by the current 
results.

\begin{question}{Darragh O'Donoghue}
Your presentation is a little more than a party political
propaganda talk. It will help now if you were to investigate why 
advocates of the long distance scale such as Feast \& Catchpole
(1997) are wrong.
\end{question}

\begin{answer}{Piotr Popowski}
It is enough if I prove that I am right, and not that they are
wrong. We have three reliable methods which consistently cluster
around a short distance to the LMC of about 45 kpc, and there is
another potentially reliable method or standard candle (Cepheids)
which gives a different result. I think that we should investigate
what is going on with Cepheids (it is the best if this is done by
Cepheid people). 
\end{answer}  

\begin{question}{David Laney} 
The statistical arguments by Feast \& Catchpole have been
demonstrated to be correct by Koen \& Laney and, using Monte
Carlo simulations, by Pont. The error bar from the Monte Carlo  
simulations is closer to 0.15 than 0.1, however.
\end{question}

\begin{answer}{Piotr Popowski}
This point has been raised by someone in the audience. No
disagreement here, so I made no comment about it.
\end{answer}

\begin{question}{Giuseppe Bono}
Two comments: 1. In a recent investigation by Romaniello et al. 
(1999) based on HST data of red clump stars, they found a distance
modulus for LMC which seems to support the long distance scale.
2. Evolutionary models suggest that the RR~Lyrae luminosity
decreases with increasing metallicity.
\end{question}

\begin{question}{G\'eza Kov\'acs}
When you talk about distance of the LMC based on RR~Lyrae stars
you should also consider RRd stars. Applying them as distance
indicators for the LMC yields $\approx$ 18.5 mag for the distance
modulus (Kov\'acs \& Walker 1998). This is in agreement with the  
Cepheid B--W distance scale.
\end{question}

\begin{answer}{Piotr Popowski}
The preliminary results of reanalysis of RRd stars from the MACHO
group (David Alves and I are involved in this work) indicate that
RRd stars' distance to the LMC is short as well.
\end{answer}

\end{document}